\documentstyle[prl,aps]{revtex}
\draft
\begin{document}
\title{On the thermodynamics of strongly correlated integrable
electron systems}
\author{Valery I. Rupasov}
\address{Department of Physics, University of Toronto, Toronto,
Ontario, Canada M5S 1A7\\
and Landau Institute for Theoretical Physics, Moscow, Russia}
\date{\today}
\maketitle
\begin{abstract}
We reexamine the Yang-Yang-Takahashi method of deriving the
thermodynamic Bethe ansatz equations which describe strongly
correlated electron systems of fundamental physical interest,
such as the Hubbard, $s-d$ exchange (Kondo) and Anderson models.
It is shown that these equations contain some additional terms
which may play an important role in the physics of the systems.
\end{abstract}


It is well known that many one-dimensional (1D) and effective
1D models describing strongly correlated electron systems of
fundamental physical interest are diagonalized exactly by
following set of the Bethe ansatz (BA) equations:
\begin{mathletters}
\begin{eqnarray}
\exp{(ik_jL)}&=&\prod_{\alpha=1}^{M}e_1(u_j-\lambda_\alpha)\\
\prod_{j=1}^{N}e_1(\lambda_\alpha-u_j)&=&
-\prod_{\beta=1}^{M}e_2(\lambda_\alpha-\lambda_\beta).
\end{eqnarray}
Here, $e_n(x)=(x+in/2)/(x-in/2)$, $k_j$ are electron
momenta, $u_j\equiv u(k_j)$, $N$ is the total number
of electrons on an interval of size $L$ and $M$ is the
number of electrons with spin ``down''. The eigenenergy
$E$ and $z$ component of the total spin of the system
$S^z$ are given by
\begin{equation}
E=\sum_{j=1}^{N}\omega_j,\;\;\;S^z=\frac{1}{2}N-M.
\end{equation}
where $\omega_j=\omega(k_j)$ are electron energies.

At $\omega(k)\sim k^2$, $u(k)\sim k$ and $\omega(k)
\sim -\cos{k}$, $u(k)\sim\sin{k}$, Eqs. (1) correspond
to a 1D electron gas with $\delta$-function interaction
\cite{G,Y} and the 1D Hubbard model \cite{LW}, respectively.
In the theory of dilute magnetic alloys \cite{TW,AFL,PS},
these equations describe the excitation spectrum of a
free host in terms of interacting Bethe excitations,
while impurity terms, omitted in Eqs. (1), account for
an elastic scattering of Bethe excitations at the impurity
site. The case of $\omega(k)=k$, $u(k)=0$ corresponds to
the $s-d$ exchange (Kondo) model \cite{W1,A}, while the
case of $\omega(k)=k$, $u(k)\sim k^2$ corresponds to the
Anderson model \cite{W2}. In addition, Eqs. (1) can be
used to study the Anderson and Kondo impurities embedded
in a host with a nonmetallic behavior of the density of
band states \cite{R1}.

The equations (1) correspond to a finite system of electrons.
A general method of deriving the thermodynamic BA equations
has been developed by Yang and Yang \cite{YY} for the case
of spinless particles, and then generalized by Takahashi
\cite{T} to the case of particles with internal degrees
of freedom.

In this Letter, we reexamine the Yang-Yang-Takahashi method
and show that the thermodynamic Bethe ansatz equations contain
some additional terms which may play an important role in the
physics of strongly correlated electron systems. These additional
terms result from a more accurate computation of a contribution
of spin degrees of freedom to a variation of the system's energy.

The structure of the BA equations shows that it is natural
to treat the system of $N-M$ electrons with spin ``up''
and $M$ electrons with spin ``down'' in terms of $N$ charge
excitations with ``rapidities'' $u_j=u(k_j)$, $j=1,\ldots,N$
and spin excitations (spin waves) with rapidities $\lambda_\alpha$,
$\alpha=1,\ldots,M$. In accordance with the Yang-Yang-Takahashi
method, to derive the thermodynamic BA equations, one needs to
note first that in the limit of a sufficiently large physical
system Eqs. (1) admit bound spin complexes in which spin
rapidities $\lambda_\alpha$ are grouped into ``strings'' of
order $n$ in the complex $\lambda$ plane,
\end{mathletters}
$$
\lambda_\alpha^{(n,j)}=\lambda_\alpha^n+\frac{i}{2}(n+1-2j),\;\;\;
j=1,\ldots,n.
$$
In addition, for many of the above-mentioned physical systems
the BA equations admit also charge complexes in which charge
excitations are bond to spin complexes \cite{TW,PS,T}.
Nevertheless, to avoid more tedious expressions we start our
analysis with the case in which charge complexes are absent.

Then, Eqs. (1) take the form of equations for charge excitation
(particle) momenta $k_j$ and rapidities of spin complexes,
$\lambda^n_\alpha$, $\alpha=1,\ldots,M_n$, where $M_n$ is the
number of complexes of order $n$ and hence $\sum_{n}nM_n=M$,
\begin{mathletters}
\begin{eqnarray}
2\pi N_j&=&k_jL+\sum_{n=1}^{\infty}\sum_{\alpha=1}^{M_n}
p_n[u(k_j)-\lambda^n_\alpha],
\\
2\pi J^n_\alpha&=&\sum_{j=1}^{N}\theta_n[\lambda^n_\alpha-u(k_j)]
-\sum_{m=1}^{\infty}\sum_{\beta=1}^{M_m}
\Theta_{nm}(\lambda^n_\alpha-\lambda^m_\beta),
\end{eqnarray}
where $p_n(x)=\pi+\theta_n(x)$, and $N_j$ and $J^n_\alpha$
are sets of quantum numbers of the system corresponding to
particles and spin complexes, respectively. Here,
$\theta_n(x)=2\arctan{(2x/n)}$ and
\end{mathletters}
$$
\Theta_{nm}(x)=(1-\delta_{nm})\theta_{|n-m|}(x)+\theta_{n+m}(x)
+2\sum_{k=1}^{{\rm min}(n,m)-1}\theta_{|n-m|+2k}(x).
$$

In the continuous limit, $L\to\infty$, $N\to\infty$,
$M\to\infty$, but $N/L$ and $M/L$ are constant, Eqs. (2)
take the form of integral equations for the ``particle''
[$\rho(k)$ and $\sigma_n(\lambda)$] and ``hole''
[$\tilde{\rho}(k)$ and $\tilde{\sigma}_n(\lambda)$] density
distributions of charge excitations and spin complexes,
respectively,
\begin{mathletters}
\begin{eqnarray}
\rho(k)+\tilde{\rho}(k)&=&\frac{1}{2\pi}
+u'(k)\sum_{n=1}^{\infty}\int d\lambda\,
a_n[u(k)-\lambda]\sigma_n(\lambda),\\
\tilde{\sigma}_n(\lambda)&=&
\int dk\,a_n[\lambda-u(k)]\rho(k)
-\sum_{m=1}^{\infty}\int d\lambda'\,A_{nm}(\lambda-\lambda')
\sigma_m(\lambda'),
\end{eqnarray}
where $a_n(x)=(2n/\pi)(n^2+4x^2)^{-1}$, $u'(k)=du/dk$, and
the matrix $A_{nm}(x)$ is given by
\end{mathletters}
$$
A_{nm}(x)=\delta_{nm}\delta(x)+(1-\delta_{nm})a_{|n-m|}(x)
+a_{n+m}(x)
+2\sum_{k=1}^{{\rm min}(n,m)-1}a_{|n-m|+2k}(x).
$$
In terms of the densities, the energy of the system is found
to be
\begin{equation}
\frac{1}{L}E=\int dk\,\omega(k)\rho(k)
\end{equation}

The number of unknown functions in Eqs. (3) is twice bigger
the number of equations. Needed additional equations for
fundamental (renormalized) energies of Bethe excitations
$$\varepsilon(k)=T\ln{\frac{\tilde{\rho}(k)}{\rho(k)}},
\;\;\;
\kappa_n(\lambda)=T\ln{\frac{\tilde{\sigma}_n(\lambda)}{\sigma_n(\lambda)}},
$$
are derived from the condition of a minimum of the thermodynamic
potential of the system
\begin{equation}
\Omega=E-TS-HS^z-AN,
\end{equation}
where $T$, $S$, $A$, and $H$ are the temperature, entropy,
chemical potential, and an external magnetic field,
respectively. The condition $\delta\Omega=0$, where the
energy's variation is given by
\begin{equation}
\frac{1}{L}\delta E=\int dk\,\omega(k)\delta\rho(k),
\end{equation}
results in
\begin{mathletters}
\begin{eqnarray}
\varepsilon(k)&=&\omega(k)-\frac{1}{2}H-A
-\sum_{n=1}^{\infty}\int d\lambda\,
a_n[u(k)-\lambda]\,F[-\kappa_n(\lambda)]\\
F[\kappa_n(\lambda)]&=&nH+\int dk\,u'(k)\,
a_n[\lambda-u(k)]\,F[-\varepsilon(k)]
\nonumber\\
&&+\sum_{m=1}^{\infty}\int d\lambda'\,
A_{nm}(\lambda-\lambda')\,F[-\kappa_m(\lambda')]
\end{eqnarray}
where $F[f(x)]\equiv T\ln{\{1+\exp{[f(x)/T]}\}}$.

The analysis of the thermodynamic BA equations (7) in the
limiting case of the Kondo system shows that the expression
(6) is incomplete. Although, in accordance with Eq. (3a) the
variation of the density distribution of particles is related
to the variation of the density distributions of spin complexes
by the relation
$$
\delta\rho(k)+\delta\tilde{\rho}(k)=u'(k)\sum_{n=1}^{\infty}
\int d\lambda\,a_n[u(k)-\lambda]\delta\sigma_n(\lambda),
$$
the latter contains only the derivative of the charge
rapidity $u'(k)$. It is easy to see that if one sets
$u'(k)=0$, as it takes place in the Kondo model, the
dependence of the energy variation on the spin degrees
of freedom completely disappears from Eq. (6), and Eq.
(7b) immediately leads, therefore, to unphysical result:
All $\kappa_n(\lambda)$ are positive, and hence spin
excitations are absent in the ground state of the system.
The expression (6) does not account thus for a total
contribution of spin degrees of freedom to a variation
of the energy of the system.

In the continuous limit for spin excitations only, Eq. (2a)
for a particle momentum,
\end{mathletters}
\begin{equation}
\frac{2\pi}{L}N_j=k_j+\sum_{n=1}^{\infty}\int
d\lambda p_n[u(k_j)-\lambda]\sigma_n(\lambda),
\end{equation}
clearly shows, however, that the momentum of a charge
excitation, and hence its energy $\omega(k_j)$, are
determined by both the quantum number of charge excitations
$N_j$ and the density distributions of spin complexes
$\sigma_n(\lambda)$ even if $u(k)=0$. Therefore, the total
variation of the system energy should be written in the form
\begin{mathletters}
\begin{equation}
\frac{1}{L}\delta E=\int dk\,\omega(k)\delta\rho(k)+
\int dk\rho(k)\delta\omega(k),
\end{equation}
where $\delta\omega(k)$ is the variation of the energy of
a particle at fixed quantum numbers of charge excitations,
$\delta N_j=0$,
\begin{equation}
\delta\omega(k)=\frac{d\omega(k)}{dk}\delta k=
-\frac{d\omega(k)}{dk}\,\frac{\sum_{n}\int d\lambda
p_n[u(k)-\lambda]\delta\sigma_n(\lambda)}
{1+2\pi u'(k)\sum_{n}\int
d\lambda a_n[\lambda-u(k)]\sigma_n(\lambda)}.
\end{equation}
Making use of Eq. (3a) for the denominator in the right-hand
side of Eq. (9b), one easily finds the final expression for the
total variation of the energy of the system,
\end{mathletters}
\begin{equation}
\frac{1}{L}\delta E=\int dk\,\omega(k)\delta\rho(k)-
\int\frac{dk}{2\pi}\frac{d\omega(k)/dk}{1+\exp{[\varepsilon(k)/T]}}
\sum_{n=1}^{\infty}\int d\lambda
p_n[u(k)-\lambda]\delta\sigma_n(\lambda).
\end{equation}
Thus, while the energy of the system is given by Eq. (4a),
its variation contains two terms. The first term corresponds
to a contribution of a variation of charge quantum numbers
at fixed spin quantum numbers of the system, while the second
one corresponds to a contribution of a variation of spin
quantum numbers at fixed charge quantum numbers.

Taking into account Eq. (10), we immediately find that the
term
\begin{equation}
-\int\frac{dk}{2\pi}\frac{d\omega(k)/dk}{1+\exp{[\varepsilon(k)/T]}}
p_n[u(k)-\lambda]
\end{equation}
should be added to the right-hand side of Eq. (7a). Then,
in the Kondo limit, where $u(k)=0$, $d\omega(k)/dk=1$, and
$$
\int\frac{dk}{2\pi}\frac{1}{1+\exp{[\varepsilon(k)/T]}}=\frac{N}{L},
$$
Eq. (7b) with the extra term (11) correctly reduces to the
standard equation describing the thermodynamics of the Kondo
model \cite{TW,AFL}.

If the Bethe spectrum of a systems contains also charge
complexes, Eq. (8) takes the form
\begin{mathletters}
\begin{equation}
\frac{2\pi}{L}N_j=k_j+\sum_{n=1}^{\infty}\int
d\lambda p_n[u(k_j)-\lambda]
\left[\sigma_n(\lambda)+\sigma'_n(\lambda)\right],
\end{equation}
where $\sigma'_n(\lambda)$ stands for the density distribution
of charge complexes of order $n$. The total variation of the
energy of the system is then found to be
\begin{equation}
\frac{1}{L}\delta E=\int dk\,\omega(k)\delta\rho(k)
-\int\frac{dk}{2\pi}\frac{d\omega(k)/dk}{1+\exp{[\varepsilon(k)/T]}}
\sum_{n=1}^{\infty}\int d\lambda
p_n[u(k)-\lambda]\delta\left[\sigma_n(\lambda)+\sigma'_n(\lambda)\right].
\end{equation}
The Yang-Yang-Takahashi method results then in the following
set of the thermodynamic Bethe ansatz equations:
\end{mathletters}
\begin{mathletters}
\begin{eqnarray}
\varepsilon(k)&=&\omega(k)-\frac{1}{2}H-A
-\sum_{n=1}^{\infty}\int d\lambda\,
a_n[u(k)-\lambda]
\left\{F[-\kappa_n(\lambda)]-F[-\xi_n(\lambda)]\right\},\\
F[\kappa_n(\lambda)]&=&nH-\int\frac{dk}{2\pi}
\frac{d\omega(k)/dk}{1+\exp{[\varepsilon(k)/T]}}
p_n[u(k)-\lambda]
+\int dk\,u'(k)\,
a_n[\lambda-u(k)]\,F[-\varepsilon(k)]
\nonumber\\
&&+\sum_{m=1}^{\infty}\int d\lambda'\,
A_{nm}(\lambda-\lambda')\,F[-\kappa_m(\lambda')],\\
F[\xi_n(\lambda)]&=&\xi^{(0)}_n(\lambda)-2nA-\int\frac{dk}{2\pi}
\frac{d\omega(k)/dk}{1+\exp{[\varepsilon(k)/T]}}
p_n[u(k)-\lambda]\nonumber\\
&&+\int dk u'(k)a_n[\lambda-u(k)]F[-\varepsilon(k)]
+\sum_{m=1}^{\infty}\int d\lambda A_{nm}(\lambda-\lambda')
F[-\xi_m(\lambda')].
\end{eqnarray}
where $\xi_n(\lambda)=T\ln{[\tilde{\sigma}'_n(\lambda)
/\sigma'_n(\lambda)]}$ are the renormalized energies of
charge complexes, while $\xi^{(0)}_n(\lambda)$ are their
bare energies, which are different for different physical
systems.

In summary, the additional terms in the thermodynamic BA
equations, derived in this Letter, should essentially
affect the thermodynamic properties of a system at finite
temperature. It is clear that the additional terms affect
essentially also the low-energy physics of such a system,
in  which particles (charge excitations unbuilt in charge
complexes) essentially contribute to the ground state of
the system, as it takes place in the Kondo and Hubbard
models. If unpaired charge excitations are absent in the
ground state of a system, as it takes place in the Anderson
model \cite{TW,PS,R2} in the absence of an external magnetic
field, the additional terms disappear from the zero-temperature
limit of Eqs. (13). In a weak magnetic field, the ground
state of the Anderson model contains also a small portion of
unpaired charge excitations, and the additional term in Eq.
(13c) should result in some small corrections to the standard
solution.

\end{mathletters}

\end{document}